\documentclass[]{spie}  
\usepackage[]{graphicx}
\usepackage{url}
\usepackage{epstopdf}  
\usepackage[labelformat=empty]{caption}
\usepackage[table]{xcolor}
\usepackage{subcaption}
\usepackage{ams math} 
\usepackage[export]{adjustbox}
\usepackage{color}
\usepackage[load=prefix,load=named]{siunitx}

\title{Deploying quantum light sources on nanosatellites I: lessons and perspectives on the optical system} 

\author{R. Chandrasekara\supit{a}, Z. Tang\supit{a}, Y.C. Tan\supit{a}, C. Cheng\supit{a}, B. Septriani\supit{a}, K. Durak\supit{a}, J.A. Grieve\supit{a} and A. Ling\supit{a}
\skiplinehalf
\supit{a}Centre for Quantum Technologies, Block S15, 3 Science Drive 2, National University of Singapore, 117543, Singapore; \\
}

\authorinfo{Further author information: Send correspondence to A. Ling, E-mail: cqtalej@nus.edu.sg, Telephone: +65 6516 2985}

\begin{document} 
  \maketitle 

\begin{abstract}
The Small Photon Entangling Quantum System is an integrated instrument where the pump, photon pair source and detectors are combined within a single optical tray and electronics package that is no larger than \SI{10}{\cm}$\times$\SI{10}{\cm}$\times$\SI{3}{\cm}. This footprint enables the instrument to be placed onboard nanosatellites or the CubeLab facility within the International Space Station. The first mission is to understand the different environmental conditions that may affect the operation of an entangled photon source in low Earth orbit. This understanding is crucial for the construction of cost-effective entanglement based experiments that utilize nanosatellite architecture. We will discuss the challenges and lessons we have learned over three years of development and testing of the integrated optical platform and review the perspectives for future advanced experiments.
\end{abstract}



\section{INTRODUCTION}
\label{sec:intro}  

Quantum Key Distrbution (QKD) comprises a family of cryptographic schemes motivated by the enhanced privacy guarantees from quantum mechanics. One of the challenges in QKD research is to extend coverage to a global network. Single photons cannot be re-amplified without destroying their quantum properties. This limits practical fiber-based QKD to about \SI{100}{\km} after which trusted nodes (which rely on traditional security mechanisms) or quantum repeaters (which are not yet demonstrated) will be necessary. 

Terrestrial free-space QKD is limited by the need for line-of-sight locations with the current record at \SI{150}{\km}. Greater distance can be achieved  by increasing the altitude of the quantum transceivers, e.g. by using long duration high altitude platforms. A compelling pathway to global QKD networks is to place core pieces of enabling technology on satellites in low Earth orbit (LEO). A number of proposals have been been published for building global quantum communication networks using satellites that host quantum light sources or detectors \cite{ursin09,ling12,morong12,scheidl13,PWQuantumSpace,jennewein14}. Efforts are under way to implement the first demonstrations. 

\begin{figure}[!h]
\centering
\begin{minipage}[t]{0.8\textwidth}
\centering
\includegraphics[width=0.8\textwidth,keepaspectratio]{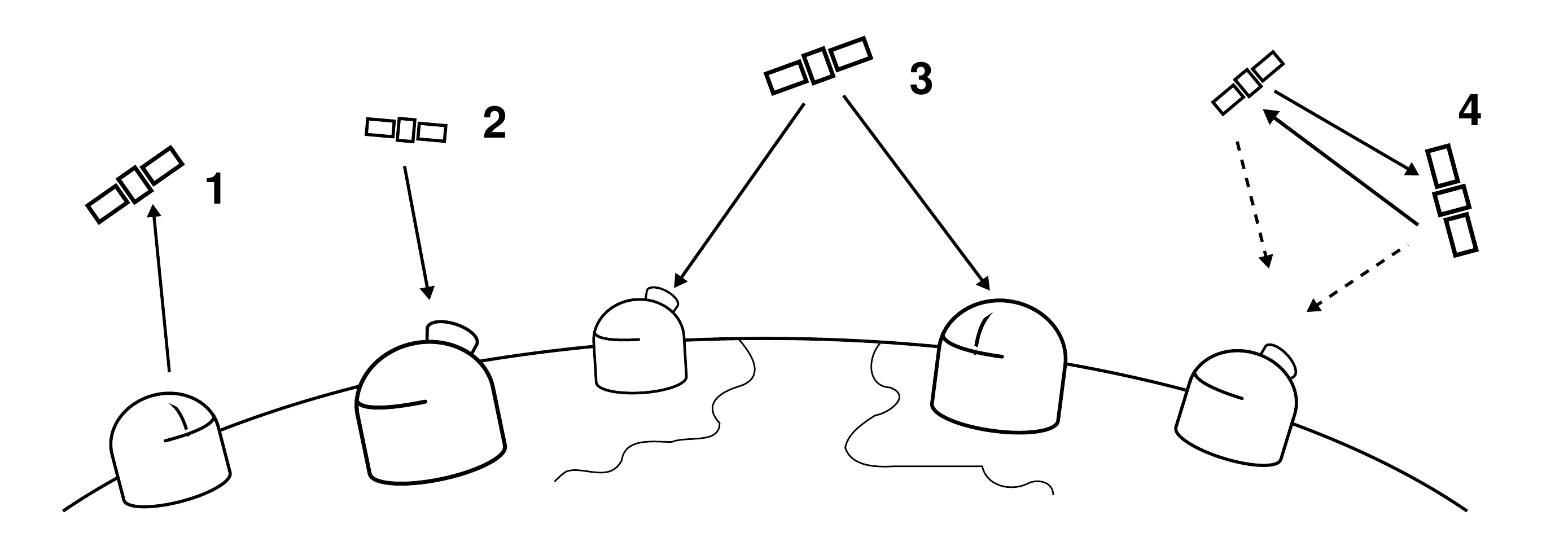}
\vspace{8pt}
\caption{{\small \textbf{Figure 1.} Possible satellite-based QKD type experiments. 1. Use of an uplink where the satellite only carries single photon detectors. 2. A downlink configuration where the satellite carries a source and detectors. 3. A platform that can beam to two ground stations simultaneously (this was the original Space-QUEST concept). 4. Inter-satellite QKD which could be the building block for a long baseline test of quantum correlations. To enable configurations 2-4 with Bell-type measurements, a source of entangled photons in space must be demonstrated. }}
\label{fig:schemes-2}
\end{minipage}
\end{figure}

In the scenario that we envision, a source of strongly correlated photons will be placed aboard cost-effective satellites, and the photons will be beamed to receiving ground stations \cite{ling12}. Although an alternative approach \cite{jennewein14} would place only single photon receivers on the satellite, our proposal reduces link loss \cite{bourgoin13} and paves the way for inter-satellite links which may be of interest when it comes to long baseline tests of quantum correlations (see Fig. \ref{fig:schemes-2}).

In the simplest implementation of space-based QKD the satellite and a ground station will establish a secret key. In a more advanced implementation, the satellite will operate as a trusted node between multiple ground stations. Although trusted nodes rely on non-quantum security assurances a space-based node is difficult to access via side-channels. For example, the reported side-channel attacks on terrestrial QKD systems \cite{gerhardt11} are difficult to implement when targeting a fast-flying node where communications can only be achieved over a solid angle of tens of pico steradians. While the pointing accuracy presents technical challenges it is worthwhile to note that optical links between satellites and ground stations already exist \cite{opals14}. The missing key enabling technology in this scenario is a working space-capable source of quantum correlated photons.

Together with collaborators we have proposed that nanosatellites (spacecraft that have a mass below \SI{50}{kg}) have a role to play in this effort. They could act as demonstrators to raise the technology readiness level of essential components and also as the final platforms that transmit and receive
single photons from ground-based stations or other satellites. In particular, we propose that nanosatellites can effectively host robust and compact sources of polarization-entangled
photon pairs, which are the workhorse for entanglement-based QKD. The decreasing cost of launching a
nanosatellite into low Earth orbit \cite{coopersmith11} has added impetus to this approach.

\begin{figure}[!h]
\centering
\begin{minipage}[t]{0.8\textwidth}
\centering
\includegraphics[width=0.8\textwidth,keepaspectratio]{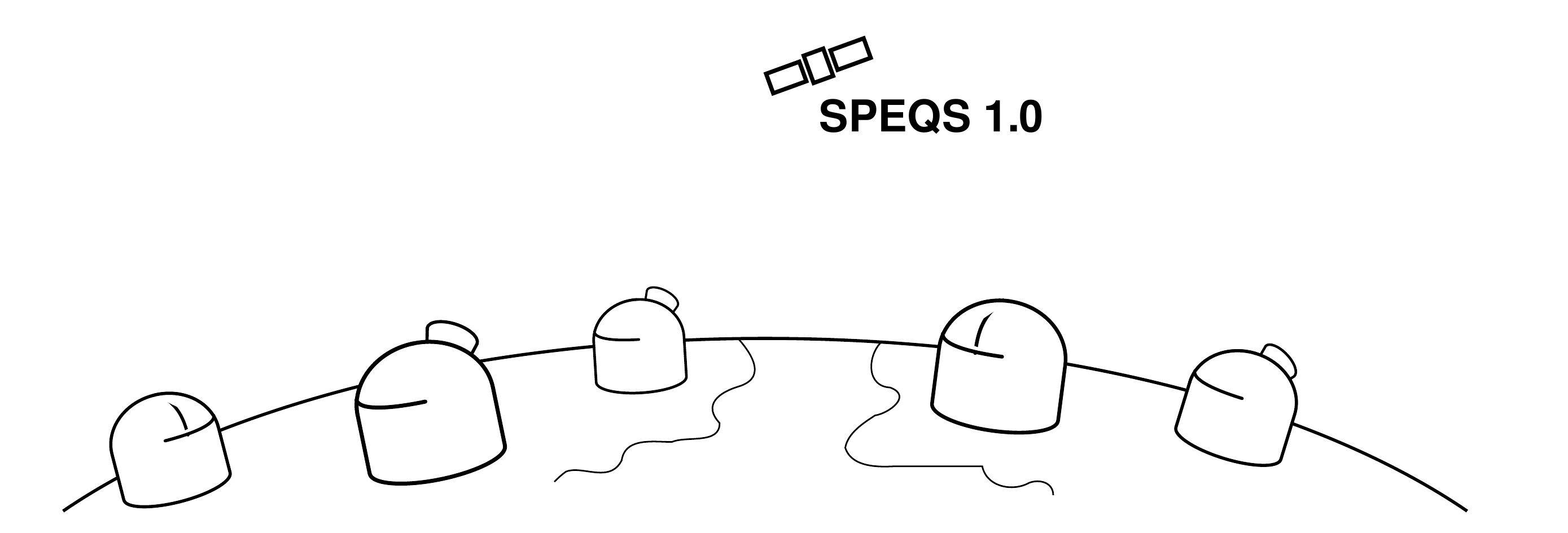}
\vspace{8pt}
\caption{{\small \textbf{Figure 2.} In our first satellite mission we aim to deploy a small and rugged correlated photon source on a (nano-)satellite to verify the ruggedness of the optical design and the operation of the supporting electronics. No optical link is needed in this configuration. }}
\label{fig:schemes-2-cqt}
\end{minipage}
\end{figure}

Our approach to this task is to take iterative steps towards a final demonstration of entanglement-based QKD from space platforms \cite{ling12}. The immediate task is to demonstrate that the basic optical design is rugged and that the control electronics are able to operate within the expected environmental envelope in LEO. To increase the chances of such a demonstration we have designed the instruments to be compatible with a widely used nanosatellite CubeSat standard \cite{woellert11}. We have designed a basic demonstration unit that minimizes size, weight and power (SWAP) requirements to increase the chance of acceptance onto a spacecraft. For this first step no pointing is necessary for the host spacecraft (see Fig. \ref{fig:schemes-2-cqt}).

An advantage of following the CubeSat standard is that it becomes also compatible with the CubeLab \cite{lumpp11} interface for operation of small science experiments on the International Space Station (ISS) should there be a requirement for ISS operation in the future \cite{ursin09,scheidl13}. The photon pair source that we are building is called the Small Photon-Entangling Quantum System 1.0 (SPEQS-1.0) and it is an integrated instrument combining low-power electronics and a rugged optical assembly. 

The goal of the SPEQS-1.0 instrument is to demonstrate that an entangled photon source works in LEO. For this to be achieved two conditions are necessary: the instrument control electronics must work, and the optical experiment must survive space operation. To build confidence in the development process it was decided to demonstrate each achievement sequentially. The first space deployment was chosen to demonstrate the electronics platform using only a basic correlated (non-entangled) photon pair source. Once the effectiveness of the electronics platform is confirmed follow-up deployments will test the optical design of the full entangled system.

In this article we will focus on the considerations that go into the SPEQS-1.0 optical design and implementation. The electronics platform has already been extensively tested and described elsewhere \cite{cheng15}. We will conclude with some comments about possible future missions.

\newpage
\section{The pros and cons of CubeSats} \label{sec:cubesat}
The primary building block of a free-flying CubeSat spacecraft is a \SI{10}{cm} cube that is a fully-functional satellite containing the main components expected within a satellite bus (e.g. solar panels, batteries, onboard computer and radio transceiver). The CubeSat architecture allows the basic cube (1U) to be stacked into larger sized spacecraft composed of multiple cubes with the 2U and 3U sizes being currently quite popular (fractional sizes are not so common). In the future it is expected that CubeSat-based spacecraft would reach 6U or larger as these will have sufficient resources to achieve more complex mission requirements.

CubeSat based spacecraft have become very popular for amateur researchers, university research groups and technology companies seeking a market niche. CubeSats are typically built from commercial-off-the-shelf (COTS) components as these satellites are usually launched into orbits with an orbital life that is measured in years or even months (although some CubeSats have been operational now for seven years \cite{bouwmeester08}). Space heritage of components is acquired from the increasing number of successful launches that share common parts. The approach of using COTS components has been highly beneficial to the SPEQS program as it has enabled the developers to focus on the design of the instrument using mature technologies that have not been formally rated for space. This avoids the use of expensive and scarce space-rated components that may have been designed to greatly outperform the requirements of a CubeSat-based mission. Space readiness is then acquired by carrying out targeted testing. This strategy has helped to reduce the necessary development time and to introduce savings enabling our small team to prepare for a space mission.

CubeSats typically ride into orbit by piggy-backing on larger satellite launches. A number of launch brokers \cite{nanoracks,spaceflight} help CubeSat developers get access to launchers or deployment opportunities via the International Space Station. Piggy-backing helps make launch costs accessible to university groups. Additionally there is a further level of savings available where each CubeSat is used to host multiple research payloads. A drawback for this approach is the low duty-cycle for experiment time as the limited power on a CubeSat results in restricted operation time for each payload.

A limitation of current CubeSat technology when it comes to meeting the requirements of high-performance optical transmission experiments is that there has been no demonstration of fine pointing on the order of a few micro-radians which is necessary for space-to-ground optical communications (crude optical communication based on optical Morse code has been demonstrated \cite{fitsat1}). Our team is in discussion with a number of groups about the feasibility of a high-performance CubeSat and we are aware by now several other groups have recently expressed interest in nanosatellite technology \cite{jennewein14}. We also propose that a source with the SPEQS form-factor is suitable even when pointing requirements necessitate a larger spacecraft.

\newpage
\section{Optical Source Considerations} \label{sec:opticalsources}
\subsection{The impact of Size, Weight and Power (SWAP) budgets} \label{sec:swap}

Candidate sources for satellite applications must satisfy the SWAP criteria while having the necessary brightness, entanglement quality and the potential for ruggedization. A large number of entangled photon sources based on various physical mechanisms have been reported in the literature but most of them cannot be considered for aerospace applications due to the SWAP requirements. The most promising class of sources are based on Spontaneous Parametric Downconversion (SPDC) and in this class only geometries relying on bulk crystals satisfy all the conditions needed for brightness, quality and ruggedization \cite{kwiat95,kwiat99,fiorentino05}. Waveguide systems are promising as future devices but are not under immediate consideration by the SPEQS-1.0 team due to the difficulties with routing pump light into the waveguide structures and the entangled photons out to the necessary detectors and optical transmitters. This also appears to be the conclusion of other teams working on space compatible light sources based on the available literature \cite{steinlechner12,steinlechner13}.

Let us first consider the SWAP restrictions. For CubeSats hosting multiple payloads the size requirement is critical. An entire entangled photon source containing also Bell-state analyzer apparatus must be contained in a volume that is approximately \SI{10}{\cm}$\times$\SI{10}{\cm}$\times$\SI{3}{\cm}. The overall mass of such an instrument should not exceed \SI{500}{\g}, while the continuous (peak) power consumption should be below \SI{2.0}{\W} (\SI{2.5}{\W}).

The SWAP restrictions lead immediately to a few key requirements for the source. The most stringent requirement is the size limit that favors collinear SPDC geometries because non-collinear sources require a more complex (and less robust) arrangement to achieve a compatible form factor. SPDC sources based on collinear geometries typically utilise quasi-phase matching conditions. Quasi-phase matching, however, often involves the use of periodically-poled crystals that require precise (within \SI{0.1}{\celsius}) temperature stabilisation. Any use of temperature stabilisation (especially with thermo-electric coolers) is challenging on a multi-payload CubeSat because of the power requirement. Furthermore, the weight restriction imposes further constraints on the size of ovens and heat sinks that can be implemented.

Two collinear geometries were considered: critical phase matching with Type-I $\beta$-Barium Borate (BBO) crystals \cite{trojek08} and a non-critical phase matching approach using Type-II periodically poled potassium titanyl phosphate (PPKTP) crystals \cite{steinlechner14}. The PPKTP source is very interesting due to the higher nonlinear coefficient, better tolerance against temperature drift and the possibility of longer crystals (with fewer compensation crystals). The BBO source however has an advantage of much lower (crystal) cost and the ability to manufacture crystals with large apertures, making alignment and collection of downconverted light much simpler. To operate with the PPKTP crystals would require beam shaping optics that were not available in the SPEQS-1.0 form-factor. After weighing the pros and cons of the two designs the BBO-based source was selected with the PPKTP system to be tested in the future.

The choice of opto-electronics parts is also a major difficulty. The major components that were selected are: Geiger-mode Avalanche Photodiodes (GM-APDs), liquid crystal polarization rotators (LCPRs), photodiodes and GaN-based laser diodes (\SI{405}{\nm}). The majority of components are obtained commercially-off-the-shelf (COTS) and come with a recommended storage and operational temperature range. Information regarding the components' survivable temperature, radiation limits, shock and random vibration tolerance are almost always unavailable. As almost all of the selected components have no space heritage the SPEQS-1.0 program has had to conduct a targeted test campaign. The results of selected tests are re-produced in section \ref{sec:tests}. 

\subsection{The SPEQS-1.0 layout} \label{sec:speqs1}
The optical set up consists of two sections: source and detection. The source section contains a laser diode, five different birefringent crystals necessary for the the production of entangled photons and the necessary optics for preparing the pump beam polarization, and redirecting excess pump light to a photodiode for monitoring and stabilizing of pump power.

\begin{figure}[!h]
\centering
\begin{minipage}[t]{0.9\textwidth}
\centering
\includegraphics[width=0.8\textwidth,keepaspectratio]{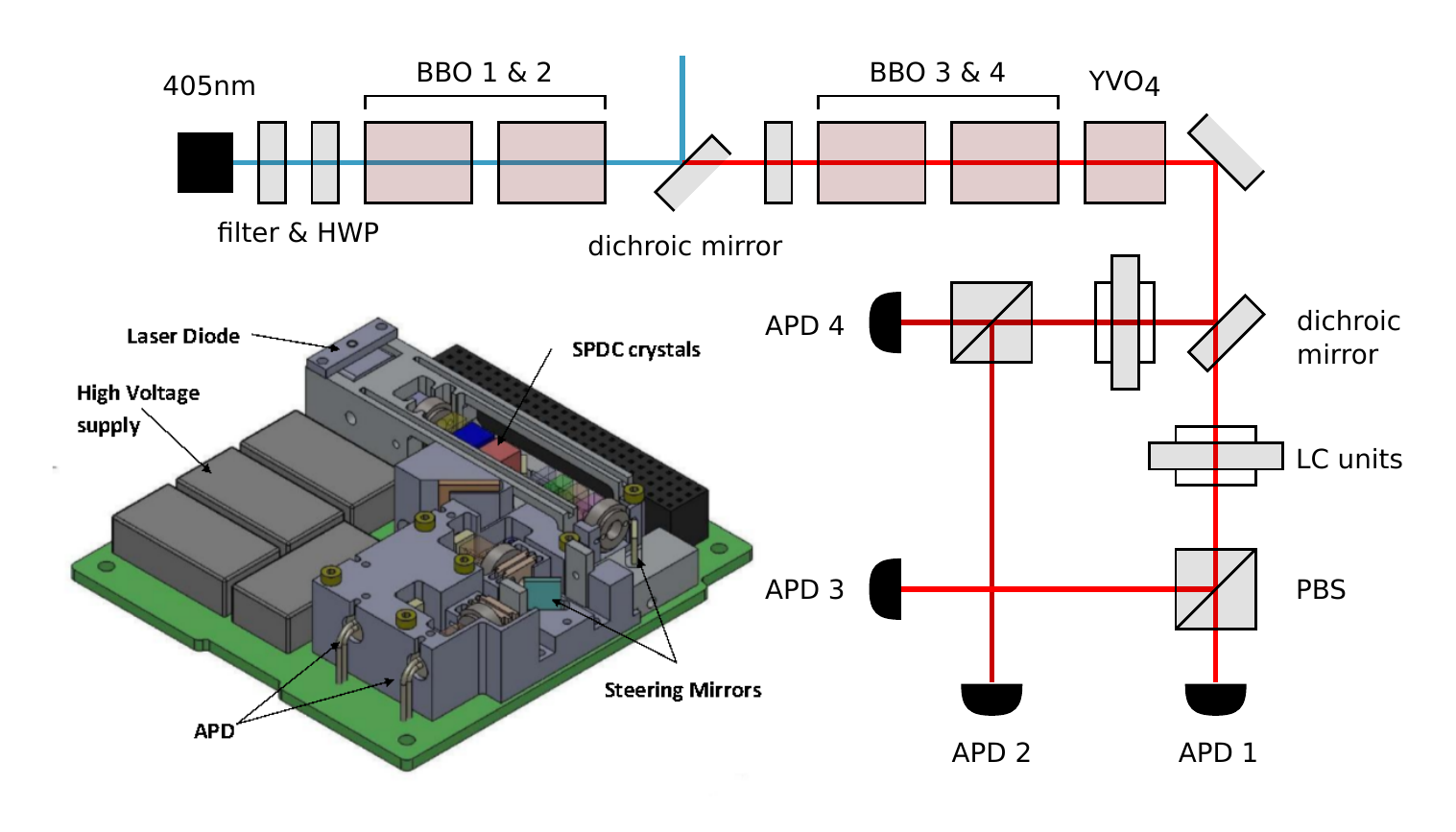}
\vspace{8pt}
\caption{{\small \textbf{Figure 3.} Optical setup of the SPEQS-1.0 instrument. The control electronics are on the reverse side of the FR4-based printed circuit board (green). The circuit board provides the mechanical interface to the spacecraft. Entangled photons are generated in $\beta$-Barium Borate (BBO1\&2). Crystals BBO3\&4 and YVO4 are used to optimize the correlations between photon pairs and to achieve entanglement. A pre-compensating YVO4 is sometimes placed just after the 405 nm laser diode (not shown). For the first mission, only BBO1 is present as correlated photons are sufficient to demonstrate the proper operation of the control electronics. Each GM-APD is provided with its own high voltage  supply (100-200\SI{}{\volt}).}}
\label{fig:source}
\end{minipage}
\end{figure}

In the detection section there is room to house four GM-APDs. By using polarization beam splitter (PBS) cubes as the polarization filters there is the option for using four detectors in total to achieve detector redundancy. In front of each PBS cube is a LCPR for the necessary polarization rotation. The layout of crystals and detectors is illustrated in Fig. \ref{fig:source}.

The entire optical unit is integrated with the electronics platform that also acts as the mechanical interface to the CubeSat. Optical components are secured to the optical unit using two component epoxy that has good mechanical strength. The cured epoxy has been tested down to $10^{-7}$\,\SI{}{\millibar} without observing outgassing \cite{manasmukherjee} and is considered suitable for the mission.

The design has no space for optical lenses. Current SPEQS-1.0 devices can process approximately $3,500$ photon pairs a second. The brightness limitation is due to the low collection efficiency of generated photon pairs (quantified using the ratio of photon pairs to single photons) because of the mismatch between pump and collection modes. For the $3,500$ observed pair events there are $200,000$ single photon events leading to a pair-to-singles ratio of $1.8\%$. The detector active area (\SI{0.5}{mm} diameter) has a $39\%$ overlap with the collimated pump mode (\SI{0.8}{mm} diameter). Assuming detector efficiency of approximately $40\%$ (and ignoring additional transmission losses through components and beam diffraction) leads to a best case collection efficiency of $2.4\%$ which is close to the observed value. Despite these losses the signal-to-noise ratio is still sufficient for a high confidence demonstration of quantum entanglement.

The original design of the optical unit was made to accomodate four detectors. Detectors 2 and 3 (see Fig. \ref{fig:source}) were designated as back-up detectors that were turned on only when necessary as they suffered from optical cross-talk due to their close proximity. The cross-talk arises from the fact that during each photo-detection event, a GM-APD will give off a flash of light due to charge carriers returning to the valence band. This detection flash is broadband and cannot be easily removed via spectral filters. Further testing has led to increased confidence in the survivability of the GM-APDs in space. In the interest of weight savings the backup detectors are not implemented in the latest instrument models.

\begin{figure}[!h]
\centering
\begin{minipage}[t]{0.88\textwidth}
\centering
\includegraphics[width=\textwidth]{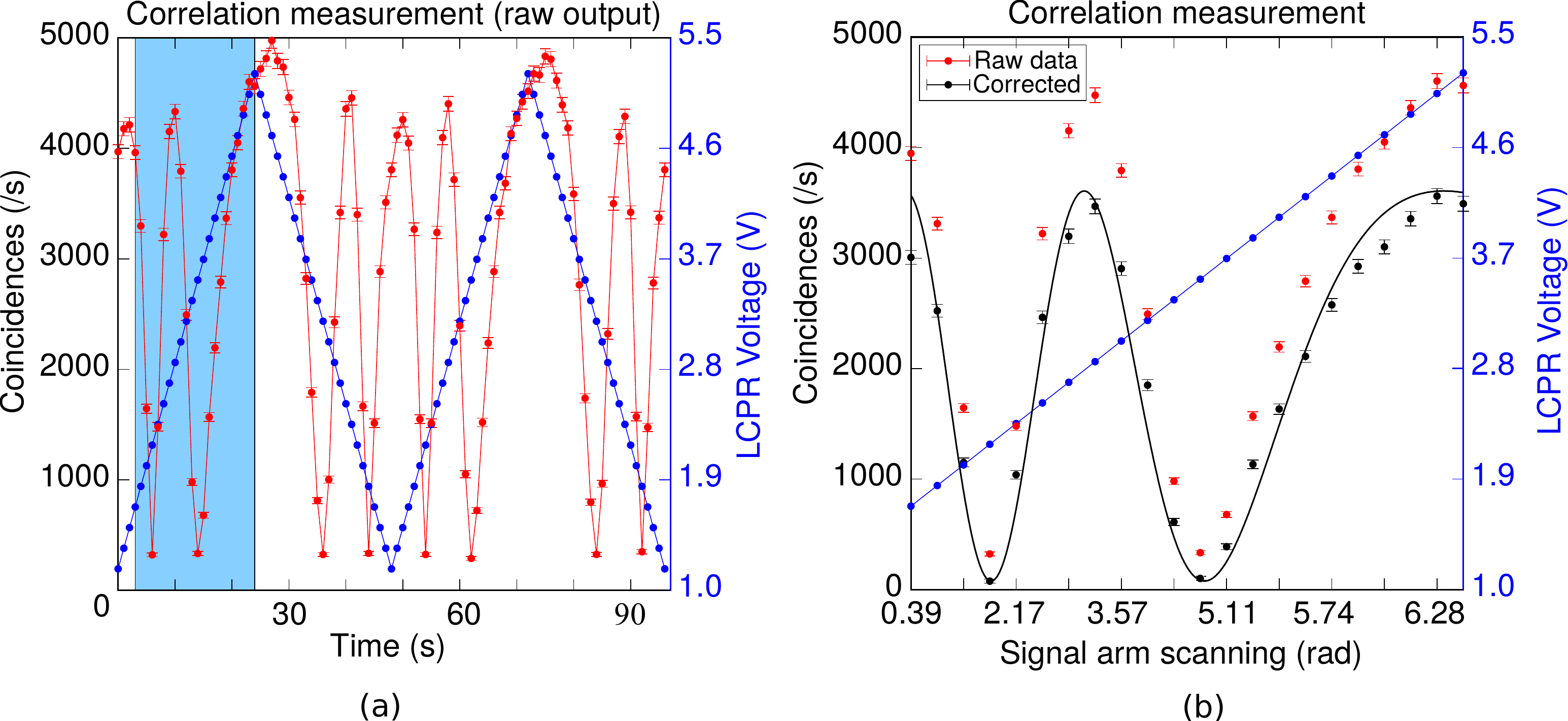}
\vspace{2pt}
\caption{{\small \textbf{Figure 4.} (a) Polarization correlations from the SPEQS instrument with one BBO crystal. Voltage to the LCPR results in a rotation of the photon polarization before it is interrogated by the polarizing beam splitter cube. (b) Polarization correlations (from the blue region of panel (a)) before and after correction for accidentals. The accidentals are underestimated in this case as there is a rate dependence that is not captured in the current model - an updated model is being prepared. }}\label{fig:rawvisi}
\end{minipage}
\end{figure}

The first mission for the SPEQS-1.0 instrument is to demonstrate the successful operation of the electronics platform \cite{cheng15} and for this reason only one BBO crystal (BBO1) is needed to generate the correlated photon pairs. The polarization correlations expected from a SPEQS device are presented in Fig. \ref{fig:rawvisi}. The qualification tests described in section \ref{sec:tests} was conducted on SPEQS instruments investigated the visibility (contrast) of the polarization correlations before and after exposure to test conditions.

\subsection{Tests} \label{sec:tests}
\begin{figure}[!h]
\centering
\begin{minipage}[t]{0.9\textwidth}
\centering
\includegraphics[width=0.95\textwidth,keepaspectratio]{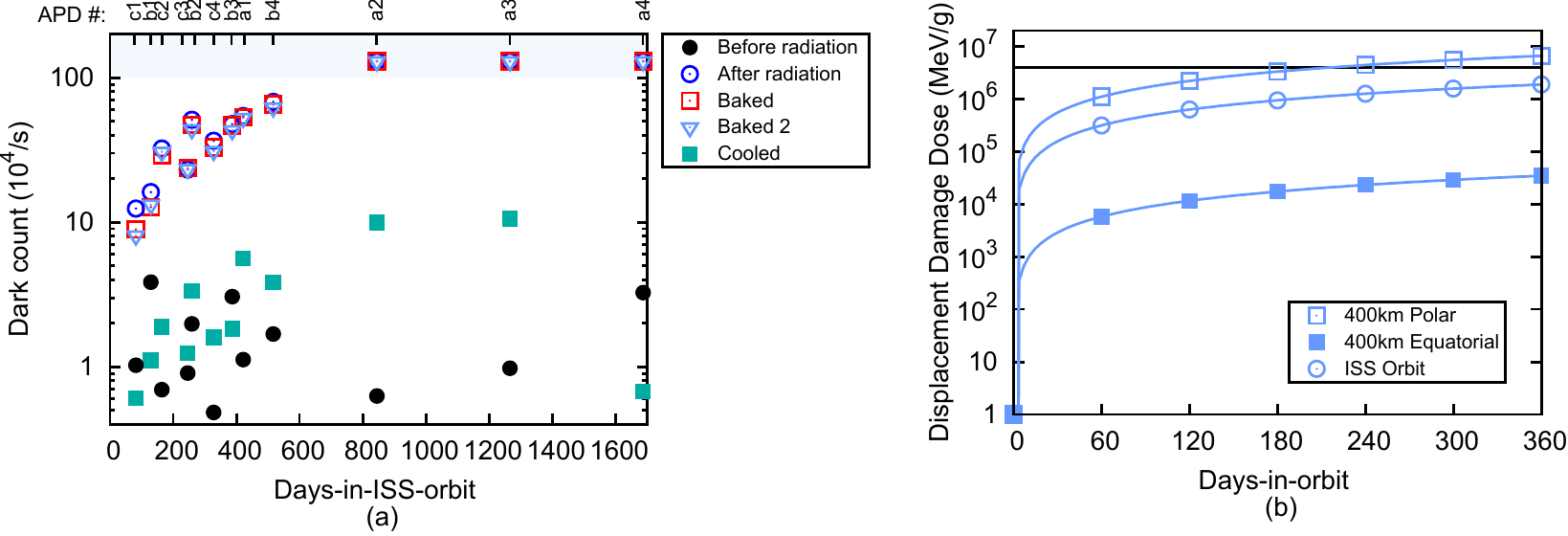}
\vspace{8pt}
\caption{{\small \textbf{Figure 5.} Results of proton irradiation exposure to GM-APDs. (a) Dark count rates of twelve GM-APDs (labeled on upper x-axis) before and after proton irradiation. Proton fluences are converted into days in ISS orbit (see (b)). After irradiation, devices were annealed at \SI{55}{\celsius} for 24 hours yielding a small but permanent drop in dark counts. Subsequently cooling the devices to \SI{-20}{\celsius} results in useful dark count rate for all devices. (b) Calculated proton Displacement Damage Dose (DDD) for three different orbits. Note the lowest radiation fluences are experienced in an equatorial orbit. }}
\label{fig:rad}
\end{minipage}
\end{figure}

Twelve GM-APDs (SAP500) were exposed to increasing doses of protons which are known to introduce strongly degrading displacement damage within Geiger-mode devices. 
The results \cite{tan13} are illustrated in Fig. \ref{fig:rad}. 
The doses of proton irradiation were converted to a Displacement Damage Dose (DDD) that can be modeled via SPENVIS software\cite{spenvis}. At different orbits the time taken to achieve a similar level of DDD depends strongly on the local radiation environment. For example orbits that take the spacecraft over the magnetic poles can induce a higher DDD compared to an orbit that is mostly along the equator. This is reflected in Fig. \ref{fig:rad} where we modeled the effect of DDD for three orbits of interest. Together the graphs strongly suggest that CubeSat based GM-APDs can remain operational over the spacecraft orbital lifetime. Furthermore if additional power were available to cool the devices several years of operational life are possible. Another device that may be sensitive to radiation damage is the LCPR. Testing this against ionizing radiation from a Co-60 source \cite{tan15} did not reveal any effects. The \SI{405}{\nm} pump diode was also exposed to proton irradiation but did not show any degradation and is consistent with observations that GaN based semiconductors are radiation tolerant \cite{hanskuiper}.

The integrated electronics platform was also exposed to radiation. In this test only the memory storage devices demonstrated sensitivity. Entire sectors of the flash memory device (M25P80 SPI Flash) were wiped out and replaced with a repeating bit pattern - presence of this repeating bit patterns can be used as evidence of radiation induced memory loss. The sectors that were erased appeared to be random across several devices. However the device remains useable as the charge pump used for writing was not degraded at low doses of irradiation \cite{raddam, c4_miyahira}. A simple mitigation strategy would be to write data to multiple physical sectors. The results of the radiation tests suggest that the vast majority of components have a very promising insensitivity to test dosage and suggest good in-orbit performance.

To test the ruggedness of the instrument a single copy was subjected to a battery of tests consisting of vacuum exposure, thermal cycling and vibration \cite{chandrasekara15_ssc}. The instrument was powered off in all tests as the objective was to demonstrate that it could withstand rough handling. During the vacuum observation the pressure was kept at $10^{-6}$\,\SI{}{\millibar} for over 24 hours. During the thermal test the instrument was subjected to a temperature cycle between $-10$\,\SI{}{\celsius} and \SI{40}{\celsius}. Each ramp duration took 50 minutes and was continued for 24 hours. The vibration test for sinusoidal and random profiles was conducted according to the Cyclone-4 launcher specifications\cite{cyclone4} (see Table \ref{table:vib}). 

The baseline visibility before the environmental tests were $93\% \pm 2\%$ and after the final (vibration) test the visibility was $92\% \pm 2\%$. It was concluded that the instrument survived testing unscathed. The next test that was conducted was a near space demonstration using a balloon which reached an altitude of \SI{35.5}{\km} \cite{tang14}. This test provided the opportunity to demonstrate the instrument in an environment where temperature and vibration/shock was not controlled and where pressure was essentially vacuum.

\begin{figure}[!h]
\centering
\begin{minipage}[t]{0.88\textwidth}
\centering
\includegraphics[width=\textwidth]{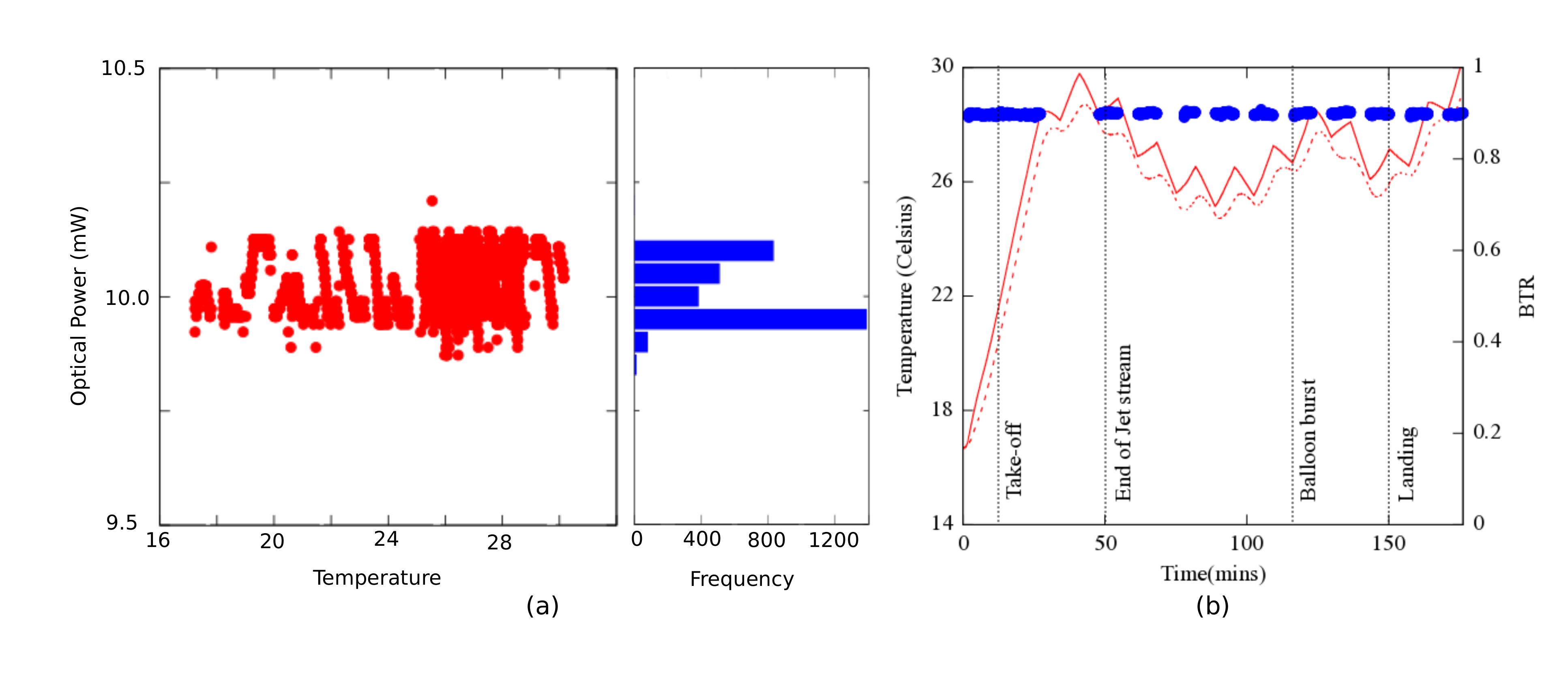}
\vspace{2pt}
\caption{{\small \textbf{Figure 6.} Performance of the laser and GM-APD stabilisation during near-space demonstration.  (a) Optical power from the laser against temperature and the corresponding histogram depicting the range of observed power values. (b) Red lines depict temperature gathered from two thermistors buried within the SPEQS optical unit; the solid line is from a sensor under the laser diode while the dashed line is from a sensor under GM-APD 1. The blue BTR points (described elsewhere \cite{cheng15}) are an indicator of excess voltage applied to the GM-APD.}}\label{fig:combopower}
\end{minipage}
\end{figure}

The temperature gradient experienced by the SPEQS instrument is approximately \SI{0.5}{\celsius} per minute and comparable to typical LEO satellite internal temperature. Fig. \ref{fig:combopower} demonstrate the temperature experienced by the SPEQS instrument during the course of the balloon flight. Both laser power and GM-APD operation were extremely reliable despite the initial rapid rise in temperature after take-off and throughout all phases including the challenging periods when the instrument was passing through the jetstream (undergoing buffeting by cold air and high velocity winds) and tumbling after the balloon had burst. The laser power remained with $0.7\%$ of nominal value while the reference GM-APD detected a constant average rate of 360,000 photons per second.

\begin{table}[h]
\renewcommand{\arraystretch}{1.3}
\caption{\small \textbf{Table 1.} Vibration test profiles.}
\label{table:vib}
\centering
\begin{tabular}{|c|c|c|}
\hline
\cellcolor{lightgray} \bfseries Description or Test Type & \cellcolor{lightgray}\bfseries{Sine Sweep} & \cellcolor{lightgray} \bfseries Random Vibration\\
\hline
X axis  & 0.5 g, 2.5 g & 7.4 g (rms)\\
\hline
Y axis & 0.5 g, 2.5 g & 7.4 g (rms)\\
\hline
Z axis & 0.5 g, 2.5 g & 7.4 g (rms)\\
\hline
Frequency (Hz) & 5-10 \SI{}{\hertz}, 10-100 \SI{}{\hertz} & 20-2000 \SI{}{\hertz}\\
\hline
\end{tabular}
\end{table}

At the conclusion of the tests efforts were made to prepare several flight models for launch on the GomX-2 spacecraft \cite{gomx2}. The flight and qualification models were subjected to additional testing as part of a complete satellite. During this test, the thermal range was actually wider (between $-20$\SI{}{\celsius} and 45\SI{}{\celsius}). After testing the visibility of the polarization correlations remained at $95\%$. The spacecraft was subsequently lost in a launch vehicle failure.

\subsection{Considerations for SPEQS-2.0} \label{sec:speqs2}
There are two improvements to the SPEQS-1.0 instrument that are being planned in the upgrade to the next generation (SPEQS-2.0) device. The first change is to swap the current cavity-based laser for a free-running device. The second change is to implement collection lenses.

The current choice of laser diode uses an external cavity and restricts the bandwidth to within a range of \SI{150}{MHz}, making it possible to dispense with the pre-compensator crystal in the source geometry. In practice the laser becomes sensitive to temperature drifts. The sensitivity is manifested as mode-hops leading to power fluctuations or phase changes that degrade entanglement quality. This requires the laser diode temperature to be stabilized to within a small range negating the original advantage of using BBO crystals that are relatively insensitive to temperature drifts. 

\begin{figure}[!h]
\centering
\begin{minipage}[t]{0.9\textwidth}
\centering
\includegraphics[width=0.95\textwidth,keepaspectratio]{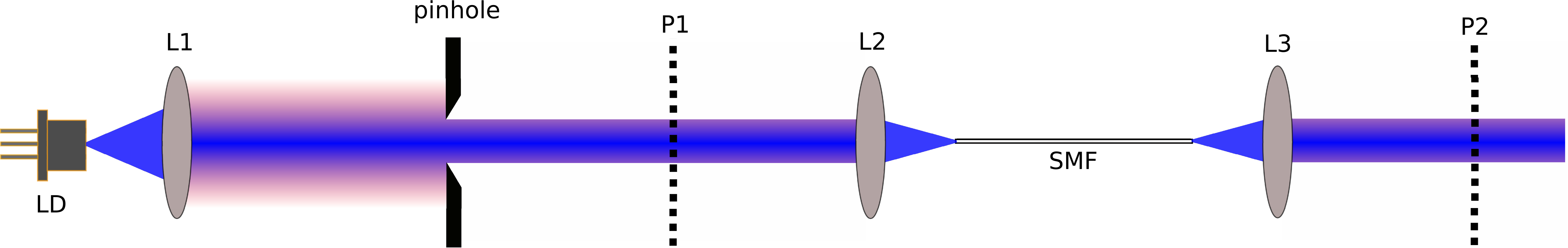}
\vspace{8pt}
\caption{{\small \textbf{Figure 7.} Spatial mode clean up for a free-running laser diode. The appropriate positioning of a pinhole after the beam is collimated can result in a mode that is compatible with a single mode fiber. This is established by demonstrating that the optical power at two positions P1 and P2 are equal (after including reflection losses). Typically after spatial filtering the beam's full-width at half-maximum is about $40\%$ of the unfiltered value. }}
\label{fig:blade}
\end{minipage}
\end{figure}

A free-running laser diode is expected to show reduced sensitivity to temperature drift. We are currently investigating the pointing stability of these devices over a temperature range, as well as designing beam shaping optics to control the pump mode within the BBO crystals. Instead of using (bulky) anamorphic prisms, a single pinhole can effectively filter the (elliptical) spatial mode of laser diodes to become compatible with single mode fibers. A pinhole that is placed in the appropriate position can transmit at least $52\%$ of light from a collimated laser diode output (power measured at P1 in Fig. \ref{fig:blade}). This filtered light can be almost completely coupled into a single mode optical fiber. Power after the single mode fiber (measured at P2) is typically about $47\%$ of the total laser diode output power. In comparison, the typical transmission of anamorphic prisms through single mode fibers is not very different from our proposed technique. We propose that for compact instruments the pinhole is a convenient method to prepare single mode beams from a laser diode.

\newpage
By implementing collection lenses the pairs-to-singles ratio of the SPEQS instrument can be increased significantly and the brightness can be sufficient for a space to ground QKD demonstration. To maximise the utility of collected SPDC photons it is desirable to collect only the emission that is compatible with the lowest order Gaussian mode (TEM$_{00}$). This can be done by collecting the light into a single-mode fiber or a series of optical baffles. Once the collected SPDC light is closely matched to the TEM$_{00}$ mode coupling to the final optical transceiver is simplified and link loss budgets over the transmission distance can be estimated. In order to effectively engineer the instrument using TEM$_{00}$ modes it is necessary to have an understanding of the pump and collection modes that optimize brightness \cite{ling08}, collection efficiency \cite{bennink10} and ruggedness. In this discussion mode size is identified as the full-width at half-maximum (FWHM) at the beam focus and it implicitly assumed that the focus of the pump and collection beams always overlap at the same point. 

To support ruggedization it is hypothesized that large mode sizes of the pump and collection beams are desirable. Any relative displacement of beams (and resulting mis-alignment) is reduced with large mode sizes. For the same reason it is hypothesized that a larger pump size will reduce the effects of spatial walk-off in the pump beam. Furthermore larger pump sizes enable maximal brightness and efficiency over a larger range of collection mode sizes. Unfortunately reports on the effects of mode size are not readily available in the literature.

An obstacle towards the development of a better understanding of beam effects on the collection of SPDC photons into the TEM$_{00}$ mode is the fact that the underlying cause of SPDC emission is vacuum fluctuation. These fluctuations are inherently multimode. However in collinear Type-I SPDC the downconverted photons exhibit a circular symmetry that matches well to the acceptance cone of a single mode fiber (or properly designed optical baffles). This circular symmetry holds even if the downconverted photons are non-degenerate and are separated by over \SI{100}{\nm} in wavelength as in the case of the SPEQS device. By working with large pump modes where the Rayleigh distance is several times the crystal length the optical rays in the pump beam can be assumed to be almost parallel thus setting up a model that follows closely the elementary phase-matching conditions. In this scenario the brightness and efficiency is determined only by the collection angle of the (single-mode) collection beam. 

\begin{figure}[!h]
\centering
\begin{minipage}[t]{0.9\textwidth}
\centering
\includegraphics[width=0.95\textwidth,keepaspectratio]{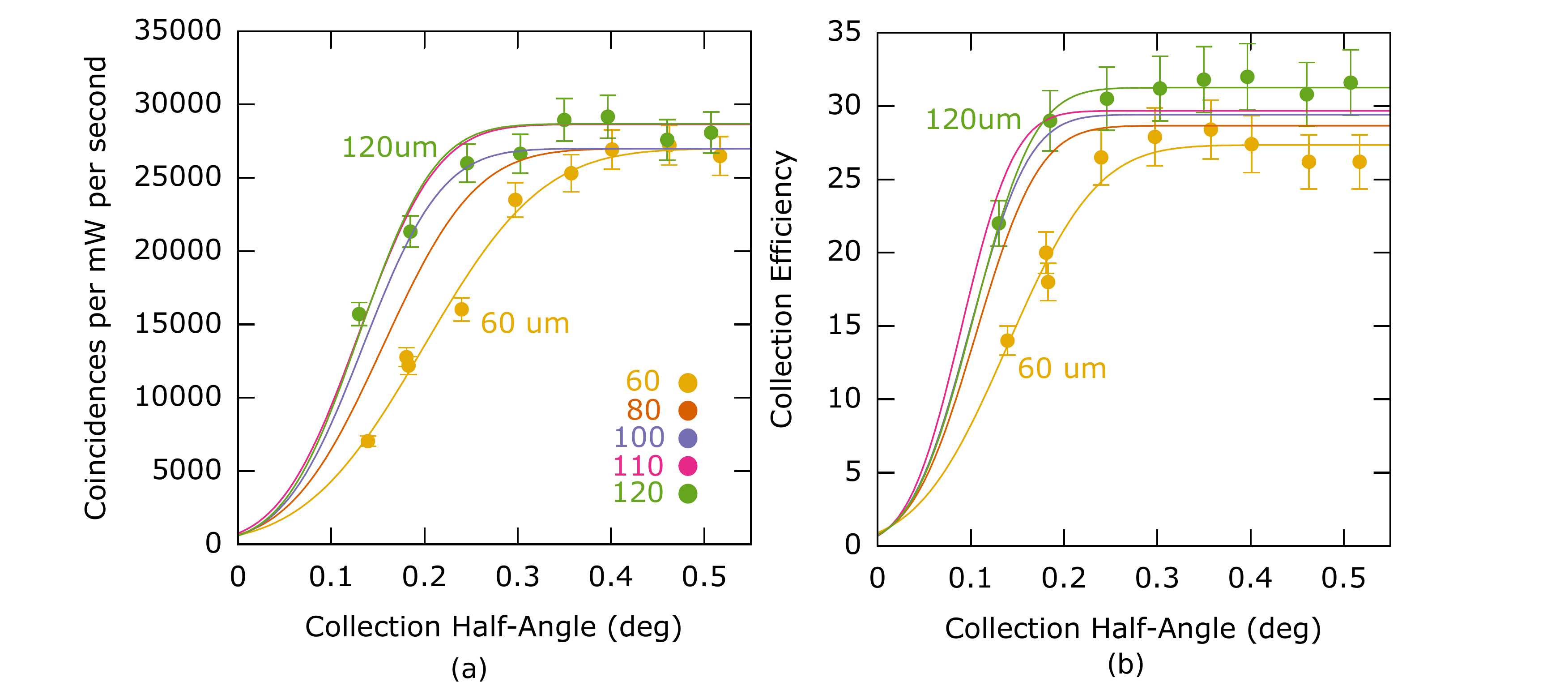}
\vspace{8pt}
\caption{{\small \textbf{Figure 8.} Effects of pump and collection beam sizes on brightness (a) and collection efficiency (b). Panel (a). The brightness rolls off more slowly when pump modes are larger. This is because the pump photons have a smaller angular spread that is carried over to the daughter photon emission pattern. The asymptotic brightness of the source remains the same. The solid lines represent error function fits for the largest and smallest pump modes. Panel (b). The collection efficiency for the corresponding data points from panel (a). The asymptotic efficiency is attained much faster (compared to the brightness). The maximum efficiency possible with the experiment is 35\% and the data suggests that collection beams were achieving an actual collection efficiency of 92\% which is in accordance with the literature \cite{trojek08}. The asymptotic values for the \SI{60}{\micro\metre} pump are slightly lower due to the use of slightly saturated detectors during data collection. }}
\label{fig:spdc}
\end{minipage}
\end{figure}

Using a Type-I SPDC geometry producing non-degenerate (\SI{785}{\nm} and \SI{837}{\nm}) photon pairs from a \SI{15}{\mm} crystal a systematic study was performed on the observed brightness and collection efficiency for different pump and collection modes (Fig. \ref{fig:spdc}). The first observation from the data is that all pump modes lead to a similar level of brightness and collection efficiency. The second observation is that the rate at which the brightness and efficiency reaches the asymptotic value does depend on the pump mode. This data set supports the the hypothesis that pump size does not affect the final brightness and collection efficiency attainable if the appropriate collection mode is used. The trend between observed brightness and efficiency with the collection angle can be attributed to the need to mode-match the angular spread of the pump and collection modes (SPDC photons have an intrinsic angular spread imposed on that of their parent photons). 

It is worthwhile to note a counter intuitive feature of the data set. All data sets reported in Fig. \ref{fig:spdc} suffer from a spatial walk-off of the pump that is approximately \SI{980}{\micro\metre}. This walk-off is substantially larger than the FWHM of the pump beams. The asymptotic collection efficiency for all the observed data, however, is $92\%$ after accounting for losses induced by components. This high value in the collection efficiency suggests that for a single crystal spatial walk-off effects are minimal. Spatial walk-off effects on entangled photon sources using pairs of crystals (as in the SPEQS geometry) will be investigated in more detail in the future.

\section{Outlook} \label{sec:outlook}
Various tests strongly suggest that the implementation of the SPEQS control electronics will be able to support photon counting experiments on board CubeSats. The current implementation of the SPEQS instrument consumes approximately \SI{1.4}{\watt} of power during normal operation while having an approximate mass of \SI{250}{\gram}.

Since 2014 the instrument has undergone two launch campaigns. The first launch campaign was with the GomX-2 spacecraft \cite{gomx2} while a second campaign is currently under way with the Galassia spacecraft \cite{luo14}. The Galassia mission is due for launch on the Polar Satellite Launch Vehicle in late 2015 with experimental data expected in the first quarter of 2016. Successful experimental data will enable the first milestone on our proposed QKD program.

The GomX-2 and Galassia missions were due to operate under very different orbital parameters. GomX-2 was supposed to have been in an orbit following the International Space Station (ISS) while Galassia is to be launched into an orbit that is nearly equatorial. The two sets of data would have enabled a comprehensive comparison of the performance of the two copies of the SPEQS instrument to understand if the orbital environment has a significant impact on the control electronics. This would be important input for future missions where choice of launchers is often restricted to polar of ISS-compatible orbits due to the presence of optical ground receivers being primarily present in the upper latitudes of the Northern hemisphere.

The SPEQS program is seeking collaborators who wish to help close this gap in orbital coverage. As the SPEQS instrument conforms strictly to the CubeSat standard devices will be made available for small satellite developers who wish to collaborate on the development of quantum technology for space. Interested colleagues are invited to contact the SPEQS team. The decision to adopt the CubeSat standard has been fruitful. By maintaining a comprehensive Interface Control Document and a straightforward Concept of Operations it has been possible to support two very different (and distantly separated) satellite groups. 

Efforts are also under way to implement the full SPEQS instrument with an entangled photon source. A test and verification campaign is being planned and should be completed by the end of 2015. A possible launch date for this full instrument is being considered for Q4 2016. The utility of sounding rockets as a proving ground for qualification testing is also being evaluated. Design work is being undertaken for the next generation device (as discussed in Section \ref{sec:speqs2}). The Critical Design Review of its supporting spacecraft (SpooQy-Sat\cite{bedington15}) is expected to be completed in 2016.

\section{Acknowledgments}
This work is supported by a grant from the National Research Foundation (NRF-CRP12-2013-02). Tan Y. C. and C. Cheng were supported by the DSO-CQT project on quantum sensors for some parts of this work. J. Grieve is supported by the MOE grant (MOE2012-T3-1-009). The authors thank R. Bedington for assistance with the text and Y. Y. Sean for some of the graphics. C. Wildfeuer facilitated the balloon flight. D. Oi has been very helpful in formulating advanced mission concepts.

\bibliographystyle{spiebib}   

\bibliography{report}   

\end{document}